\documentclass[twocolumn,showpacs,preprintnumbers,amsmath,amssymb]{revtex4}

\usepackage{graphicx}
\usepackage{dcolumn}
\usepackage{bm}

\newcommand{\be}{\begin{equation}}
\newcommand{\ee}{\end{equation}}
\newcommand{\bea}{\begin{eqnarray}}
\newcommand{\eea}{\end{eqnarray}}
\newcommand{\eps}{\varepsilon}

\newcommand{\half}{{\scriptstyle{\frac{1}{2}}}}
\newcommand{\real}{\relax{\rm I\kern-.18em R}}
\newcommand{\integer}{\relax{\rm I\kern-.18em N}}
\newcommand{\Tr}{{\rm Tr}}

\begin{document}
\title{On the stability of Dirac sheet configurations}
\author{E.-M. Ilgenfritz}
 \altaffiliation[Presently at ]{Instituut-Lorentz for Theoretical Physics, 
University of Leiden, PO Box 9506, NL-2300 RA Leiden, The Netherlands}
\author{M. M\"uller-Preussker }
 \affiliation{Humboldt-Universit\"at zu Berlin, Institut f\"ur Physik, 
Newtonstr. 15, D-12489 Berlin, Germany}
\author{ B.V. Martemyanov}
 \altaffiliation[Presently at ]{Instituut-Lorentz for Theoretical Physics, 
University of Leiden, PO Box 9506, NL-2300 RA Leiden, The Netherlands}
 \affiliation{ Institute for Theoretical and Experimental Physics, 
B. Cheremushkinskaya 25, 117259 Moscow, Russia}
\author{Pierre van Baal }
 \affiliation{Instituut-Lorentz for Theoretical Physics, 
University of Leiden,\\ PO Box 9506, NL-2300 RA Leiden, The Netherlands }
\date{}

\begin{abstract}
Using cooling for $SU(2)$ lattice configurations, purely Abelian constant 
magnetic field configurations were left over after the annihilation of 
constituents that formed metastable $Q=0$ configurations.
These so-called Dirac sheet configurations were found to be stable 
if emerging from the confined phase, close to the deconfinement phase 
transition, provided their Polyakov loop was sufficiently non-trivial. 
Here we show how this is related to the notion of marginal stability of 
the appropriate constant magnetic field configurations. We find a perfect 
agreement between the analytic prediction for the dependence of stability 
on the value of the Polyakov loop (the holonomy) in a finite volume and 
the numerical results studied on a finite lattice in the context of the 
Dirac sheet configurations.

\end{abstract}


\maketitle

\vspace{-5mm}
\section{Introduction}

\vspace{-3mm}
In lattice gauge theory cooling is used to remove the high frequency
fluctuations to be left with classical solutions~\cite{Cool,Garc}. 
This allows one to extract the underlying topological content of the gauge 
field configurations and determine to what extent instantons have a role to 
play. It is known that when using the ordinary Wilson action, the lattice 
artefacts are such that one can further lower the action by reducing the 
size of the instantons (whereas in the continuum the classical action does 
not depend on the size). Ultimately the instanton falls through the lattice 
and in general one relaxes to the trivial minimum with zero action. However, 
at finite temperature, when the Polyakov loop away from the instanton is 
non-trivial, the relevant instanton (called a caloron) actually consists of 
$n$ constituents for SU($n$)~\cite{KvB,LLu}. 
These can be shown to be 't Hooft-Polyakov (BPS) monopoles when identifying 
$A_0$ with the (adjoint) Higgs field. From the Euclidean four-dimensional 
point of view, due to the self-duality of the gauge field, these are dyons 
with their magnetic charge equal to their electric charge, with overall 
electric and magnetic neutrality.

Under cooling in the confined phase, due to the discreteness artefacts of the 
Wilson action, these constituents will attract and approach each other. When 
they are no longer visible as separate entities, the solutions behave like  
ordinary instantons localized in space and time. The distance between the 
constituents is (for SU(2)) given by $\pi\rho^2/b$, where $b$ is the inverse 
temperature (the period in the euclidean time direction). 
Another possibility is 
the annihilation of dyons and antidyons left over from different caloron and 
anticaloron solutions. As a result, with an action near the one-instanton 
action, a metastable configuration can be either a dyon-dyon pair that shrinks 
and falls through the lattice or a dyon-antidyon pair that finally 
annihilates~\cite{IMMPSV}. Sometimes this annihilation process leaves behind 
a constant Abelian magnetic field, which subsequently turns out to be stable 
or unstable under further cooling~\cite{DS}, strongly correlated to the 
asymptotic value of the Polyakov loop (the holonomy) which has been acquired 
in this stage of cooling. In the deconfined phase no dyonic structure was 
observable under cooling. The Polyakov loop remains always close to its 
trivial value but quasi-constant magnetic field configurations were seen to 
emerge as well, although they never happened to be stable. In this paper we 
present an explanation for these observations.

\vspace{-5mm}
\section{Constant magnetic fields}

\vspace{-3mm}
It is well-know that Abelian constant magnetic fields are embedded solutions 
of the (non-Abelian) equations of motion. They tend to be unstable, due to the
self-coupling of the gauge fields~\cite{Savv}, which formed the basis for the 
studies of the so-called Copenhagen vacuum picture~\cite{Cope}. 

In the four-dimensional context a constant field is stable if it is 
self-dual~\cite{Leut,Tho2}. Here we will be interested in the degenerate 
case with magnetic, but no electric flux and periodic boundary conditions 
(the general case allows for center fluxes, but requires twisted boundary
conditions~\cite{Tho1}). For $SU(2)$ these gauge fields are Abelian and 
there is the freedom of adding to it a constant Abelian vector potential,
which does not change the field strength, $F_{\mu\nu}=\pi i \tau_3 
n_{\mu\nu}/(L_\mu L_\nu)$. This field strength is unique up to a constant 
gauge rotation, and $n_{\mu\nu}$ is an integer (even in the case of periodic 
boundary conditions) antisymmetric tensor, fixed by flux quantization. 
In the degenerate case $n_{\mu\nu}$ has two non-zero eigenvectors, and 
computing the gauge-invariant Polyakov-loop observables in this subspace 
it is easily seen that no translation invariance holds. Adding a constant 
Abelian vector potential can consequently be absorbed by a translation and 
can therefore not affect the fluctuation spectrum. But in this degenerate 
case there are also two zero eigenvectors, and the vector potential is 
invariant under translations in this subspace. Its Polyakov loops label 
the gauge invariant parameters on which the fluctuation spectrum do depend! 

It had been found~\cite{Marg,Lat95} that on a symmetric torus there was one 
class of constant magnetic field solutions that for a certain range of values 
of the Polyakov loop were stable. This example, involving the smallest possible 
non-zero magnetic field, possesses non-trivial center flux and requires twisted 
boundary conditions. Therefore it could not explain the findings of 
Ref.~\cite{DS}. However, at finite temperature involving a non-symmetric box, 
more room exists to obtain stable constant magnetic field solutions. 

\vspace{-5mm}
\section{The fluctuation spectrum}

\vspace{-3mm}
For $SU(2)$ all constant curvature solutions in a finite box have been 
classified. Also the spectrum of fluctuations has been calculated~\cite{Baa2}. 
For the ``charged'' isospin components, in the subspace of non-zero 
eigenvectors of $n_{\mu\nu}$, the problem is equivalent to that of Landau 
levels. The eigenfunctions are described by $\Theta$ functions to incorporate 
the boundary conditions. In the subspace of zero eigenvectors one simply has 
plane waves, with properly discretized momenta. These momenta are, however, 
shifted due to the constant vector potential which determines the Polyakov 
loops in this subspace, thereby obviously modifying the fluctuation spectrum. 
The ``neutral'' isospin component is described by ordinary plane waves.

The following gauge field for $SU(2)$ gives the most general solution with 
constant field strength on a torus~\cite{Lat95}
\be
A_\nu(x)=\half i(-\pi n_{\nu\mu}x_\mu/L_\mu+C_\nu)\tau_3/L_\nu \; .
\ee
It is periodic up to the gauge transformation
\be
A_\nu(x+\hat\mu L_\mu)=\Omega_\mu(x)(A_\nu(x)+\partial_\nu)\Omega_\mu^{-1}(x)
\label{eq:twist}
\ee
where $\hat\mu$ is the unit vector in the $\mu$ direction and 
\be
\Omega_\mu(x)=\exp(\half i\pi x_\nu n_{\nu\mu}\tau_3/L_\nu) \; .
\ee
With $n_{\mu\nu}$ even, these Abelian boundary conditions are, however, gauge 
equivalent (in general by a non-Abelian gauge transformation) to periodic 
boundary conditions (as long as $Q=0$). 
Following Ref.~\cite{DS} we assume $L_0=L_t=b$, 
$L_1=L_2=L_3=L_s$. The data can in all cases be interpreted in terms
of a (nearly) constant magnetic field with $n_{0\nu}=-n_{\nu 0}=0$ and 
$\vec m=(0,0,2)$, where $m_i=\half\eps_{ijk}n_{jk}$. Therefore we compute
the fluctuation eigenvalues for this case (compare Ref.~\cite{Marg,Lat95,Baa2})
\bea  
\lambda_\pm & = & 4\pi(2n+1\pm2)/L_s^2    \nonumber \\
            &   &   +(2\pi p+C_3)^2/L_s^2 \nonumber \\
            &   &   +(2\pi q+C_0)^2/L_t^2 \; , \\
\lambda_0   & = & (2\pi k_\mu/L_\mu)^2 \; . \nonumber 
\eea
The multiplicities are $4$ for $\lambda_\pm$ and 2 for $\lambda_0$,
with all quantum numbers ($n,p,q,k_\mu$) integer (but $n\geq 0$).

As argued above the spectrum depends on the constant Abelian gauge field 
described by the constants $C_0$ and $C_3$. These are only defined modulo 
$2\pi$, as a shift over $2\pi$ is related to a gauge transformation 
that shifts the relevant momenta by one unit. The Polyakov-loop observables 
are given by
\be
P_\mu=\half\Tr\exp(iC_\mu\tau_3/2)=\cos(C_\mu/2) \; ,\quad\mu=0,3 \; .
\ee
Note that these are anti-periodic under a shift over $2\pi$, whereas the 
fluctuation spectrum is periodic. This is simply because the fluctuations 
involve fields in the adjoint representation, whereas the Polyakov loop is 
in the fundamental representation. Indeed $P_0^2$ and $P_3^2$, relevant for 
the adjoint representation, are periodic under a shift of $C_0$ and $C_3$ 
over $2\pi$.

{}From the lattice data it is clear that $P_3=1$, and we can put $C_3=0$,
as well as $n=p=q=0$ (we may restrict $|C_0|\leq\pi$) to find the lowest 
eigenvalue $\lambda_-=-4\pi/L_s^2+C_0^2/L_t^2$ to be negative unless
the Polyakov loop is sufficiently non-trivial ($P_0=\pm1$ being associated
to a trivial Polyakov loop). The lowest eigenvalue is positive when
\bea
 L_t/L_s & < & \sqrt{\pi}/2 \nonumber \\
   & &                 \label{eq:cond} \\
 |P_0| & = & \cos(C_0/2)<\cos(\sqrt{\pi} L_t/L_s) \; . \nonumber
\eea
We see that these conditions cannot be satisfied when $L_t=L_s$ and
the finite temperature situation ($b=L_t<L_s$) is essential for providing
the opportunity of stability. 

This stability was called marginal, because one can change $C_0$ without 
changing the classical action. Thus nothing prevents us to bring $C_0$ close 
to 0, where the lowest eigenvalue $\lambda_-$ turns negative. Under the 
cooling there is no reason for $C_0$ to change, as one can easily show that 
the degeneracy of the action as a function of $C_0$ survives on the lattice. 
This then explains the stability of these constant gauge field configurations, 
provided the two conditions of Eq.~(\ref{eq:cond}) are satisfied.

\vspace{-5mm}
\section{Comparison with lattice data}

\vspace{-3mm}
In Ref.~\cite{DS} $SU(2)$ gauge theory in four-dimensional Euclidean space 
was considered on an asymmetric lattice with periodic boundary conditions 
in all four directions. The respective ensembles of configurations have been 
created by heat-bath Monte Carlo using the standard Wilson plaquette action.
The lattice size was $N_s^3 \times N_t$ with the spatial extension 
$N_s = 8,10,12,16,20$ and with temporal extension $N_t=4$, {\it i.e.} $b=4a$
and $L_s=aN_s$ with $a$ the lattice spacing). 
For $N_t=4$ the model is known to undergo the deconfinement phase transition 
at the critical coupling $\beta_c \simeq 2.299$~\cite{Bielefeld}. 
In Ref.~\cite{DS} two ensembles with $\beta_1 = 2.2 < \beta_c$ and 
$\beta_2 = 2.4 > \beta_c$ were generated.

The equilibrium field configurations in both ensembles have been cooled by 
iterative minimization of the Wilson action 
with the focus at the structure of selfdual caloron solutions. In addition, 
Dirac sheet (DS) events were observed at the very last stages of cooling, 
applying a stopping criterion which selects action plateaux in the interval 
$S \le 0.6~S_{\mathrm{inst}}$. In confinement, approxmately 7 \% (at $N_s=8,10,12$), 
5 \% ($N_s=16$) and 3 \% ($N_s=20$) of equilibrium configurations 
have turned into these purely magnetic configurations, whereas 
in the deconfinement phase the yield was 5 $\cdots$ 18 \% \cite{DS}.
The action values were found close to $(N_t/N_s)S_{\mathrm{inst}}$ 
characteristic for constant Abelian magnetic flux~\cite{Mitrjushkin}
of size $4\pi$ periodically closed along one of the spatial directions.
\begin{figure*}
  \centering
    \includegraphics[height=5cm]{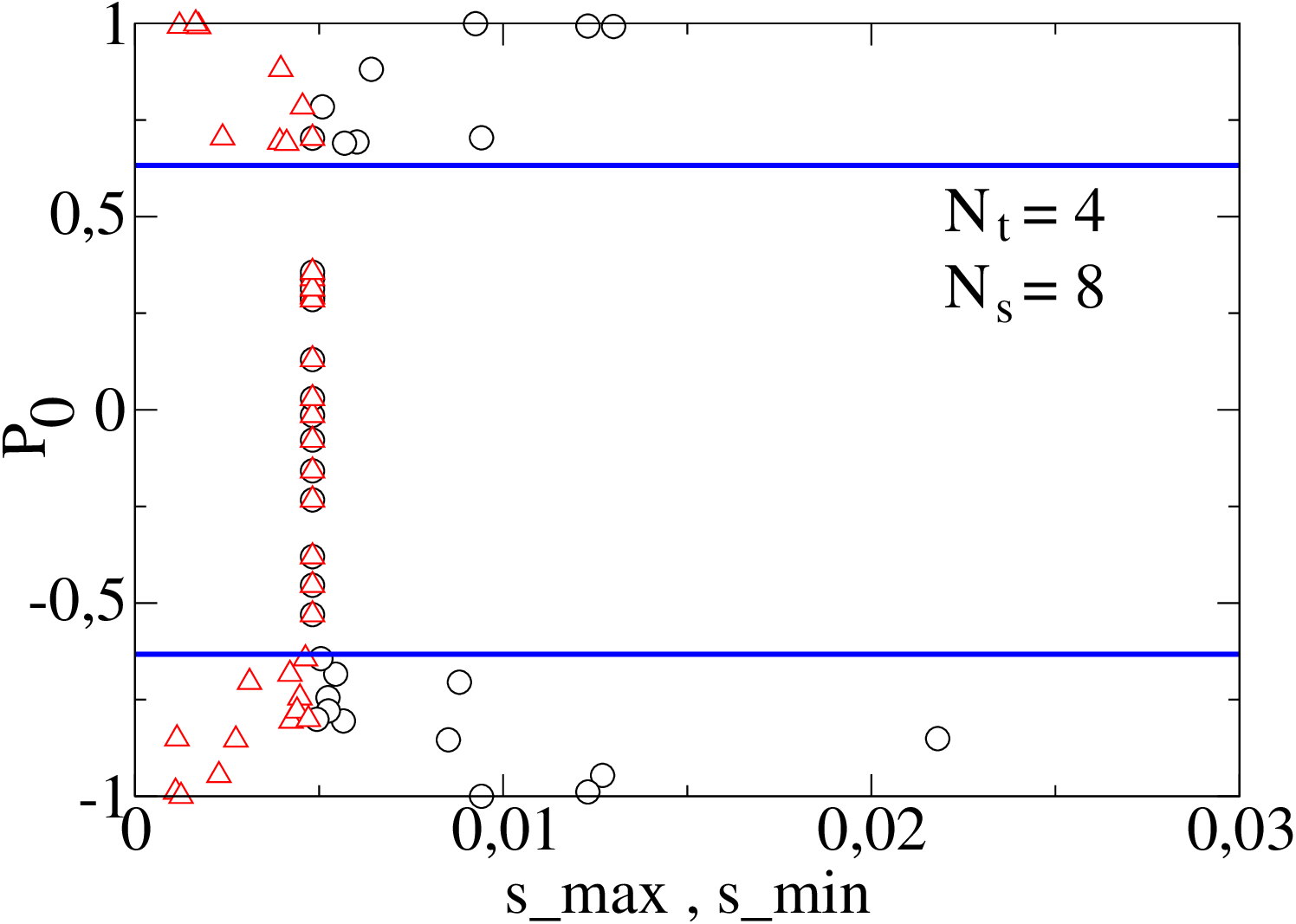}\qquad%
    \includegraphics[height=5cm]{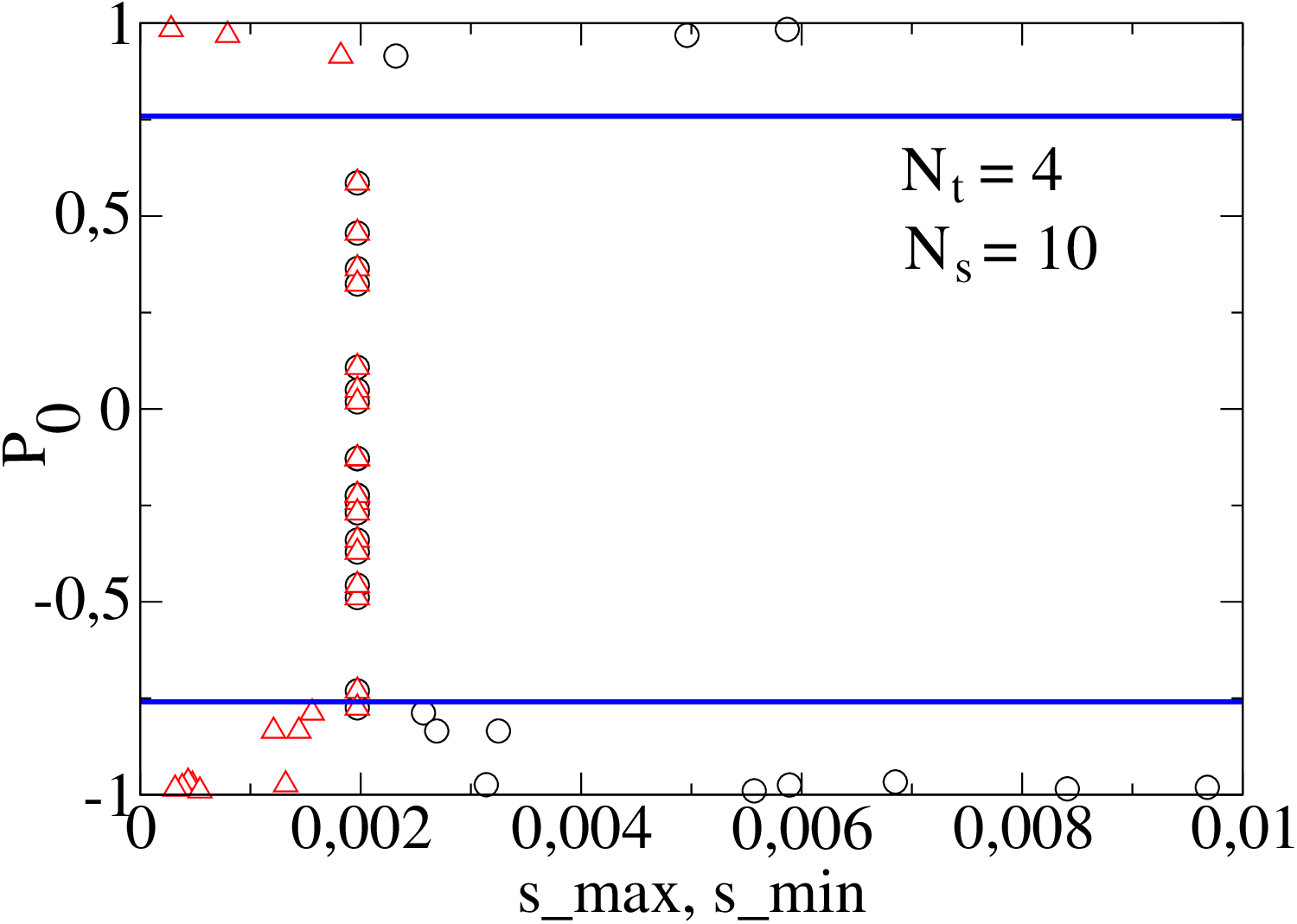}
    
\vspace{1.0cm}

    \includegraphics[height=5cm]{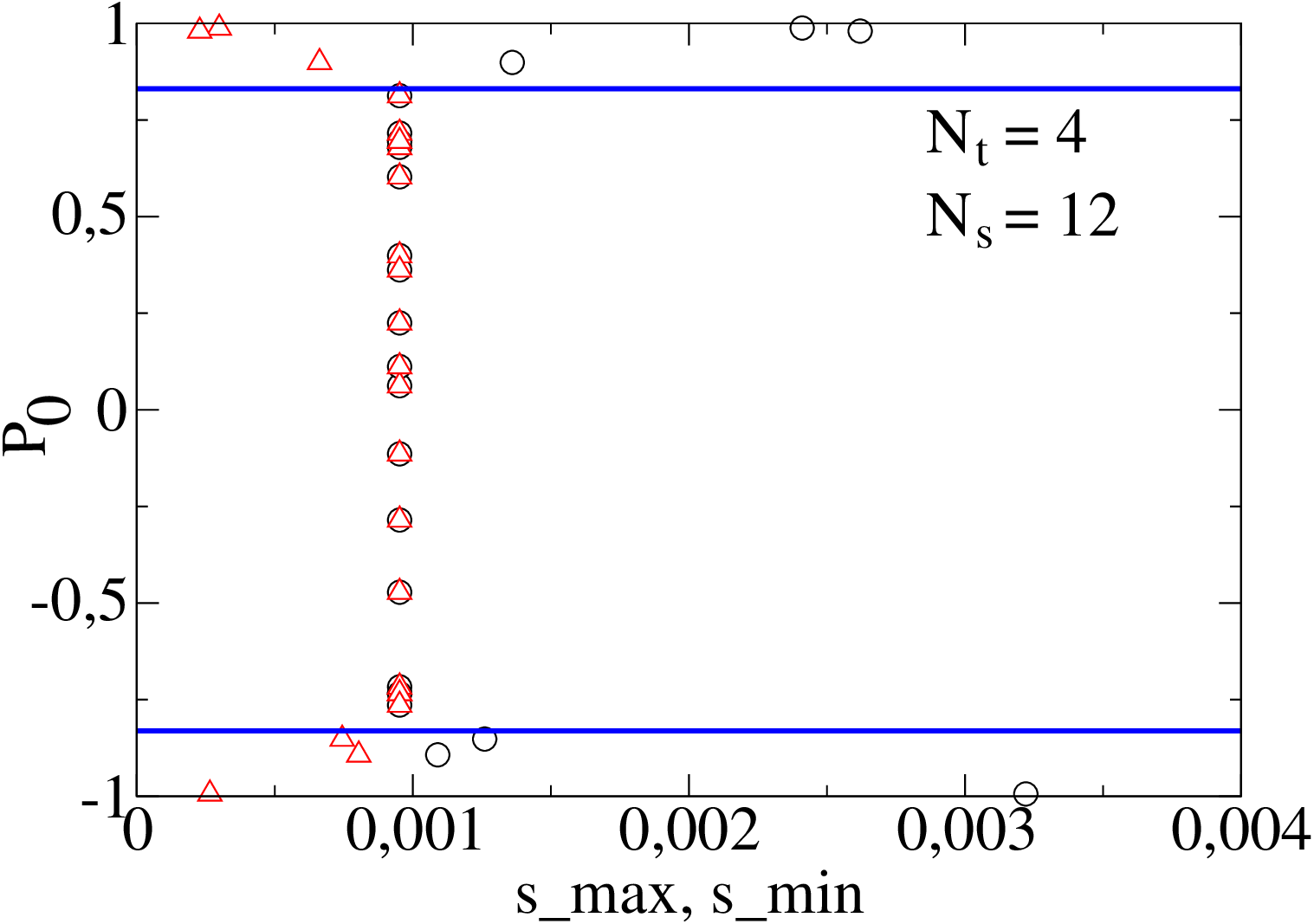}\qquad%
    \includegraphics[height=5cm]{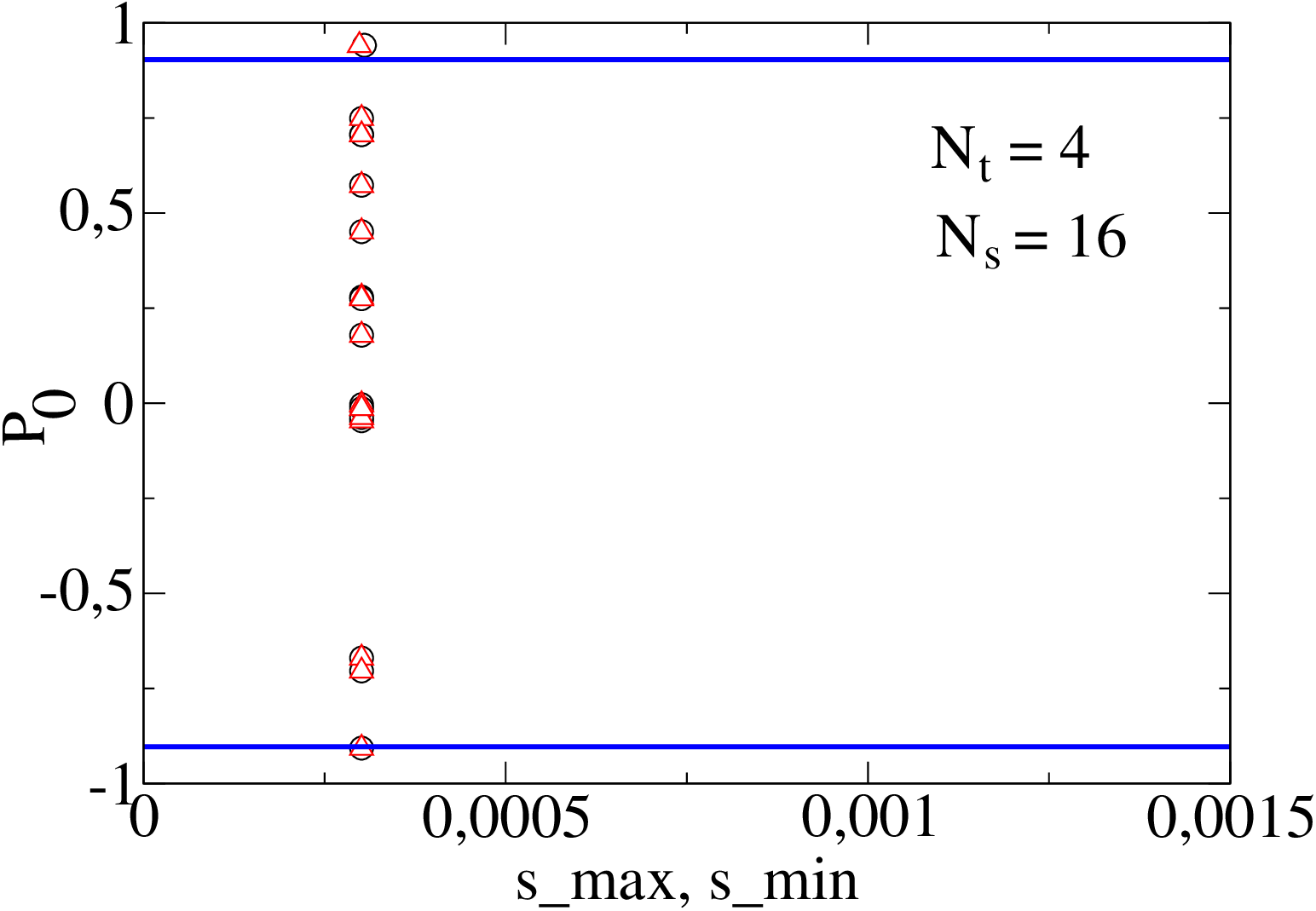}
    
\vspace{1.0cm}

    \includegraphics[height=5cm]{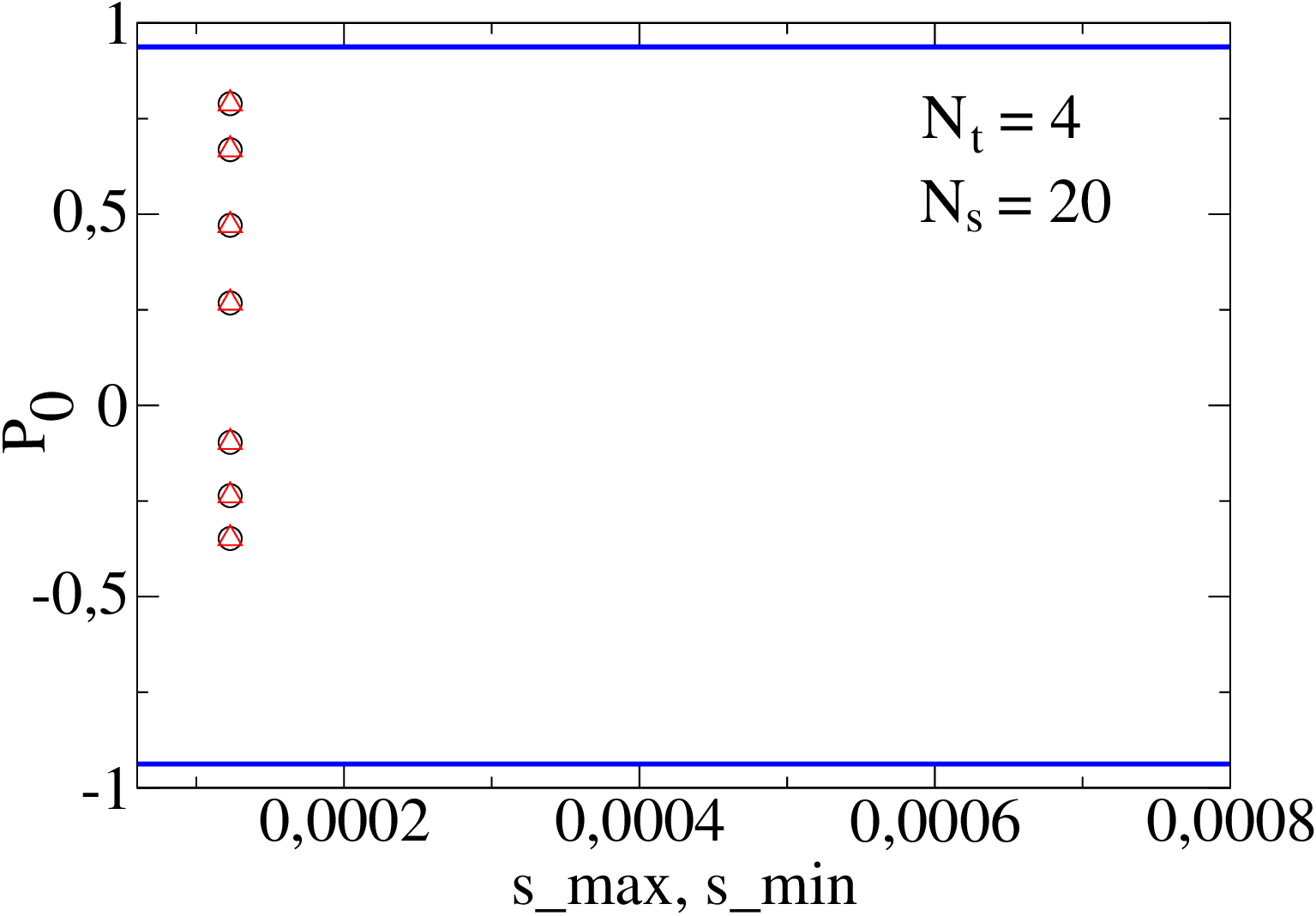}

    \caption{Correlation in the confined phase of $SU(2)$ LGT
     between the holonomy $P_0$ and the (in)stability of Dirac 
     sheets which is indicated through the remaining inhomogeneity 
     of the action density, $s_{max} \neq s_{min}$ (represented 
     by circles and triangles, respectively). The limiting values 
     of the holonomy, $|P_0|=\cos(\sqrt{\pi}\frac{N_t}{N_s})$, 
     beyond which constant magnetic fluxes become unstable are 
     indicated by horizontal lines. The temporal size of the 
     lattices is $N_t=4$, and the DS events are shown for different 
     spatial lattice size $N_s=8$, $10$, $12$, $16$, $20$.} 
\label{corrLstab}
\end{figure*}
Although the action showed the same dependence on the lattice extensions $N_t$ 
and $N_s$, 
supporting the common interpretation as (almost) homogeneous magnetic flux,
the configurations were unstable when derived from the deconfined phase and 
partly stable in the case of the confinement phase. In the case of confinement, 
the issue of stability {\it vs.} instability was strongly correlated to the 
value of the temporal Polyakov line (holonomy) $P_0$. This is shown in Figure 
\ref{corrLstab}. It presents a set of scatterplots (each for another spatial 
size $N_s$) where each DS event is characterized by two entries: ($s_{min}$, $P_0$) 
and ($s_{max}$, $P_0$). The values $s_{min}$ and $s_{max}$ express the action 
density at sites where it is minimal and maximal, respectively. If these 
values differ, the configuration is bound to decay to the trivial vacuum. 
Provided the holonomy remains sufficiently far from trivial, we find only 
DS events which consist of a highly homogeneous Abelian magnetic flux 
signaled by $s_{min}=s_{max}$. This case is tantamont to absolute stability 
under further cooling.  
In contrast to this, when the holonomy was close to the trivial one 
($P \approx \pm 1$) the Abelian magnetic fluxes happened not to be homogeneous 
($s_{min}i \neq s_{max}$) and proved to be unstable under further cooling. 
The critical value of the holonomy, $|P_0|=\cos(\sqrt{\pi} L_t/L_s)$, limiting 
the region of stability as given by the second condition in Eq.~(\ref{eq:cond}), 
is marked in Figure \ref{corrLstab} by horizontal lines. {\em No deviations} 
from the predicted (in)stability are seen.

\vspace{-5mm}
\section{Conclusions}

\vspace{-3mm}
Purely Abelian constant magnetic field configurations were observed~\cite{DS}, 
randomly emerging from the process of cooling-down equilibrium lattice fields 
representing the confined and deconfined phases of $SU(2)$ gluodynamics. 
In the confined phase they were found to be absolutely stable provided their 
Polyakov loop was sufficiently non-trivial. We have shown here that this fact
is related to the notion of marginal stability of the appropriate constant 
magnetic field configurations. We have found perfect agreement between the 
analytically predicted dependence of stability on the value of the Polyakov 
loop (the holonomy) for the set of spatial lattice sizes that were studied in
Ref.~\cite{DS} and the numerical observations made there, separating stable 
from unstable Dirac sheet configurations. 
The dependence on the geometry of the effect we found makes us believe we 
are dealing with a finite volume artefact. Nevertheless, it demonstrates that 
the influence of a background constant $A_0$ (as manifested by the Polyakov loop) 
on the dynamics of the gauge field should not be ignored. The physical significance 
lies in the fact that the Polyakov loop is the order parameter for the 
confinement/deconfinement phase transition. 
A similar conclusion can be drawn from the caloron solutions with non-trivial 
holonomy.

\vspace{-5mm}
\section*{Acknowledgements}

\vspace{-5mm}
This work was  supported by FOM and by RFBR-DFG, grant 03-02-04016. 
Two of us (E.-M. I. and B. V. M.) gratefully appreciate 
the hospitality experienced at the Instituut-Lorentz of Leiden University.


\vspace{-3mm}

\begin{thebibliography}{99}

\bibitem{Cool}
B. Berg, Phys. Lett. {\bf B 104}, 475 (1981);
E.-M. Ilgenfritz, M. L. Laursen, G. Schierholz, M. M\"uller-Preussker 
and H. Schiller, Nucl. Phys. {\bf B 268}, 693 (1986); 
J. Hoek, M. Teper and J. Waterhouse, Nucl. Phys. {\bf B 288}, 589 (1987).
\bibitem{Garc}
M. Garc\'{\i}a P\'erez, A. Gonz\'alez-Arroyo, J. Snippe and P. van Baal, 
Nucl. Phys. {\bf B 413}, 535 (1994) [hep-lat/9309009].
\bibitem{KvB}
T. C. Kraan and P. van Baal, Phys. Lett. {\bf B 428}, 268 (1998) 
[hep-th/9802049]; Nucl. Phys. {\bf B 533}, 627 (1998) [hep-th/9805168]. 
\bibitem{LLu}
K. Lee and C. Lu, Phys. Rev. {\bf D 58}, 025011 (1998) [hep-th/9802108].
\bibitem{IMMPSV}
E.-M.~Ilgenfritz, B.V.~Martemyanov, M.~M\"uller-Preussker, S.~Shcheredin, 
and A.I.~Veselov, Phys. Rev. {\bf D 66}, 074503 (2002) [hep-lat/0206004].
\bibitem{DS}
E.-M. Ilgenfritz, B.V. Martemyanov, M. M\"uller-Preussker and A.I. Veselov, 
''On Dirac sheet configurations of SU(2) lattice fields'', hep-lat/0310030.
\bibitem{Savv}
G. K. Savvidy, Phys. Lett. {\bf B 71}, 133 (1977).
\bibitem{Cope}
P. Olesen, Phys. Scr. {\bf 23}, 1000 (1981).
\bibitem{Leut}
L. S. Brown and W. I. Weisberger, Nucl. Phys. {\bf B 157}, 285 (1979);
H. Leutwyler, Nucl. Phys. {\bf B 179}, 129 (1981); 
P. Schwab, Phys. Lett. {\bf B 109}, 47 (1982).
\bibitem{Tho2}
G. 't Hooft, Comm. Math. Phys. {\bf 81}, 267 (1981).
\bibitem{Tho1}
G. 't Hooft, Nucl. Phys. {\bf B 153}, 141 (1979).
\bibitem{Marg}
M. Garc\'{i}a P\'erez and P. van Baal, 
Nucl. Phys. {\bf B 429}, 451 (1994) [hep-lat/9403026].
\bibitem{Lat95}
P. van Baal, Nucl. Phys. {\bf B} (Proc.Suppl.) {\bf 47}, 326 (1996) 
[hep-lat/9508019]. 
\bibitem{Baa2}
P. van Baal, Comm. Math. Phys. {\bf 94}, 397 (1984). 
\bibitem{Bielefeld}
J.~Engels, J.~Fingberg, and V.~K.~Mitrjushkin, Phys. Lett. {\bf B 298}, 154 
(1993); \\ J.~Engels, S.~Mashkevich, T.~Scheideler, and G.~Zinovev, 
Phys. Lett. {\bf B 365}, 219 (1996).
\bibitem{Mitrjushkin}
V. K. Mitrjushkin, Phys. Lett. {\bf B 389}, 713 (1996).
\end{thebibliography}
\end{document}